\documentclass[reprint,twocolumn,superscriptaddress,aps]{revtex4-2}

\usepackage[utf8]{inputenc}
\usepackage[british]{babel}

\usepackage{amsfonts}
\usepackage{amssymb}
\usepackage{amsmath}
\usepackage{amsthm}

\usepackage{bbold}
\usepackage{bm}
\usepackage{graphicx}
\usepackage{xcolor}
\usepackage[hidelinks]{hyperref}
\usepackage{multirow}
\usepackage{float}
\usepackage{array}
\usepackage{csquotes}

\usepackage{booktabs}
\usepackage{orcidlink}

\usepackage{natbib}
\definecolor{b}{rgb}{0.4, 0.7, 1.}
\usepackage{tikz}
\def \minwidth {3.}
\def \textwid {2.5}
\usetikzlibrary{shapes.geometric, arrows.meta, tikzmark, calc, positioning}
\tikzstyle{ell} = [ellipse, minimum width=3.5cm, minimum height = 1.cm, draw = black, text centered, text width=3cm]
\tikzstyle{rec} = [rectangle, minimum width=\minwidth cm, minimum height = 1.2cm, draw = black, text centered, text width=\textwid cm, rounded corners]
\tikzstyle{wrec} = [rectangle, minimum width=0.1 cm, minimum height = 1.2cm, draw = white, text width=0.1 cm]
\tikzstyle{link} = [->, >={Stealth[width=1.7mm, length=1.7mm]}]
\tikzstyle{nlink} = [thick]
\tikzstyle{circ} = [circle, minimum width=0.5 cm, draw = black, text width=0.5 cm, text centered]

\graphicspath{ {./figures/} }

\usepackage{indentfirst}
\newcolumntype{M}[1]{>{\centering\arraybackslash}m{#1}}

\newcommand{\tab}{Tab.}

\begin{document}

\title{Using time series to identify strongly-lensed gravitational waves with deep learning}
\author{Arthur Offermans\,\orcidlink{0000-0002-8313-5976}}
\affiliation{Department of Physics and Astronomy, KU Leuven, B-3001 Leuven, Belgium}
\affiliation{Leuven Gravity Institute, KU Leuven, Celestijnenlaan 200D box 2415, 3001 Leuven, Belgium}

\author{Tjonnie G. F. Li\,\orcidlink{0000-0003-4297-7365}}
\affiliation{Department of Physics and Astronomy, KU Leuven, B-3001 Leuven, Belgium}
\affiliation{Leuven Gravity Institute, KU Leuven, Celestijnenlaan 200D box 2415, 3001 Leuven, Belgium}
\affiliation{KU Leuven, Department of Electrical Engineering (ESAT), STADIUS Center for Dynamical Systems, Signal Processing and Data Analytics, B-3001 Leuven, Belgium }

\begin{abstract}
    The presence of a massive body between the Earth and a gravitational-wave source will produce the so-called gravitational lensing effect. In the case of strong lensing, it leads to the observation of multiple deformed copies of the initial wave. Machine-learning (ML) models have been proposed for identifying these copies much faster than optimal Bayesian methods, as will be needed with the detection rate of next-generation detector. Most of these ML models are based on a time-frequency representation of the data that discards the phase information. We introduce a neural network that directly uses the time series data to retain the phase, limit the pre-processing time and keep a one-dimensional input. We show that our model is more efficient than the base model used on time-frequency maps at any False Alarm Rate (FPR), up to $\sim 5$ times more for an FPR of $10^{-4}$. We also show that it is not significantly impacted by the choice of waveform model, by lensing-induced phase shifts and by reasonable errors on the merger time that induce a misalignment of the waves in the input.
\end{abstract}

\maketitle

\section{\label{sec:intro}Introduction}
In 2015, a gravitational wave was directly detected for the first time~\cite{Abbott_2016}. Since then, a bit less than 100 events have been detected~\cite{GWTC1,GWTC2,GWTC3} by the LVK collaboration. A specific event, GW170817~\cite{Abbott_2017,Abbott_2019_param}, stands out for the numerous studies and tests that resulted from its detection (\textit{e.g.}~\cite{Abbott_2019_TGR,Drout_2017}). It is also the first multi-messenger (electromagnetic + gravitational waves) detection. All these detections are only the beginning of the gravitational-wave astronomy, as next-generation detector are predicted to detect up to about $10^6$ binary black hole mergers per year~\cite{Maggiore_2020}, and other sources, including multi-messenger candidates, are yet to be found.

As the number of detections increases, we expect to find instances of events that have a much smaller detection rate than usual events. Gravitational lensing is one of these rarer instances. 
This effect occurs when a wave passes by a massive objects as it travels towards the Earth. This object, called a lens, deflects and distorts the wave. In some cases, such as galaxy or galaxy cluster lenses for example, it results in the observation of several copies of the same waveform with different amplitudes, times of arrival and possibly phases~\cite{Takahashi_2003,Deguchi_1986,Nakamura_1998,Ohanian_1974}. The case with multiple images is referred to as strong lensing and is the focus of this paper. 
The detection of lensed waves has many scientific applications, including characterization of the lens, search for dark matter, intermediate-mass and primordial black holes~\cite{Tambalo_2023,Oguri_2020,Lai_2018,Diego_2020,Basak_2022,Caliskan_2023}, cosmology~\cite{Li_2019, Sereno_2011, Liao_2017}, precise source localization~\cite{Hannuksela_2020, Uronen_2024, Wempe_2024}, and tests of general relativity (GR) and constraints on beyond GR effect~\cite{Ezquiaga_2020, Goyal_2021b,Goyal_2023,Narola_2024,Baker_2017,Mukherjee_2020}. Searches for lensing signatures in gravitational-wave data have been conducted~\cite{Hannuksela_2019,Abbott_2021,Abbott_2023,Dai_2020,Li_2023,Liu_2021,Singer_2019,Janquart_2023,McIsaac_2020}, reporting no conclusive evidence for lensed events despite some arguments in favour of the presence of lensed events in the data~\cite{Broadhurst_2018, Diego_2021}. 

The main idea to identify strongly-lensed gravitational waves relies on the fact that the lensed waves have the same intrinsic parameters, such as masses, mass ratio and spins, since they correspond to the same single source. The source position, on the other hand, is different because of the deflection angle caused by the lens. However, since the sky localization accuracy of gravitational wave detectors is not high enough to resolve this deflection angle, the two signals should have overlapping skymaps. The estimated parameters of lensed events should thus be consistent with each other. Bayesian analysis provides us with a way to compare the estimated parameters properly. The parameter posteriors computed by a parameter estimation analysis should then overlap. Bayesian hypothesis test is thus used to determine if the data are better explained within the lensed or unlensed scenario (\textit{e.g.}~\cite{Janquart_2022b,Cheung_2023}). The parameter estimation required for this technique as well as the technique itself can become time-consuming and it would thus not be feasible to keep up with the increasing detection rate. It would also be very inefficient to apply the full method on each possible pairs, considering that the proportion of lensed events is predicted to be small~\cite{Xu_2022,Oguri_18,Wierda_21,Mukherjee_21,Li_18,Chen_24}. This will become an even more serious issue for next-generation gravitational-wave detectors, where the number of pairs that would need to be analyzed is $O(10^{10})$. Several analysis pipelines have been proposed~\cite{Janquart_2022a,Ezquiaga_2023,Lo_2023,Haris_2018,chakraborty_24a,chakraborty_24b}, but they may not be fast enough for the detection rate of next-generation detectors.

Motivated by this issue, machine learning models have been proposed~\cite{Goyal_2021,Magare_2024}, as they are known to allow for fast predictions once the training is done. The proposed models are based on the time-frequency representations of the data. This representation has however two main drawbacks. The first is that it requires a significant part of pre-processing and transforms a one-dimensional input into a two-dimensional space, \textit{i.e.} generally an input of larger size (depending on the time and frequency resolutions). Both will thus tend to make the model slower. The second is that the phase is lost through the transformation to the time-frequency plane. This has the advantage that the model will not be influenced by phase shifts caused by lensing, so this will not worsen its performance. It is however also a drawback in the sense that this phase (and phase shift) could be a very important information to determine whether two events are strongly lensed or not.\\

In this paper, we propose a deep learning model that uses the time series directly to allow for fast inference and to retain the phase information. We use data generated without any model for the lens nor lensing rates to study the potential of the time domain representation based on the waveform only and excluding the potential influence of any model-dependent factor. Although the phase may be an obstacle for the network to identify identical waveforms, we show that our model can nevertheless distinguish lensed pairs from unlensed ones better than the base existing model trained on time-frequency maps. We also analyze three factors that are susceptible to affect the performance of the network given the time domain representation that we use: the misalignment of waves in the input, the waveform approximant and the phase shift induced by lensing. 

The paper is organized as follows. First, we present our data sets and how lensing is taken into account in those data. Then, we present the neural network model that we used as well as the the training strategy. After that, we present the results of our tests on our own data sets and compare with~\cite{Goyal_2021} on their own data. Finally, we summarize our results and conclude.

\section{\label{sec:method}Method}

\subsection{\label{sec:Data}Data}

Contrary to~\cite{Goyal_2021} and~\cite{Magare_2024}, our model is used on time series data directly and does not require transformation to a time-frequency plane. The advantage of using the time series is that the phase of the signal is preserved. We expect it to improve the classification as more information is provided. It could also be a crucial discriminating factor between a lensed and an unlensed pair, as the coalescence phase of two unlensed events can be different, and the observed phase difference may not be consistent with the time delay between the events.
The last advantage is the lower computational cost. Indeed, almost no pre-processing of the data is needed and the representation is one-dimensional. As a result, for a same duration of signal the 1D model will be faster than the 2D one. A drawback is that the time resolution is higher and hence more sensitive to the merger time uncertainty used to align the different events. Although it does impact the performance of our model for large uncertainties, we will show that this effect is much less significant than the effect of the signal-to-noise ratio (SNR).\\

To build our data sets for training, validation and test of the network, we generate a set of waveforms whose parameters are drawn randomly according to their distribution, which are shown in \tab~\ref{tab:parameters}, and using the \texttt{IMRPhenomXPHM}~\cite{Pratten_2021} approximant. Our data sets thus contain only waveforms that are not yet projected onto any detector, and not pairs of events.

\setlength{\extrarowheight}{1.0ex}
\begin{table}[H]
    \centering
    \begin{tabular}{M{0.32\linewidth} M{0.30\linewidth} M{0.30\linewidth}}
        Parameter & Distribution & Range\\
        \toprule[0.2ex]
         Chirp mass & Uniform & $[6, 50]M_\odot$ \\[1ex]
         Mass ratio & Uniform & $[0.1, 1]^*$\\[1ex]
         Coalescence phase & Uniform & $[0, 2\pi]$\\[1ex]
         Inclination angle & Sine & $[0, \pi]$\\[1ex]
         Right ascension & Uniform & $[0, 2\pi]$\\[1ex]
         Declination & Cosine & $[-\pi/2, \pi/2]$\\[1ex]
         Polarization angle & Uniform & $[0, \pi]$\\[1ex]
         \bottomrule[0.2ex]
    \end{tabular}
    \footnotesize{\\[1ex]$^*$For a given chirp mass, the minimum mass ratio is chosen such that the obtained minimum black hole mass is above $5M_\odot$ and maximum below $100 M_\odot$.}
    \caption{Distribution considered for each parameter. The geocentric time of merger is drawn uniformly in a period of half a year.}
    \label{tab:parameters}
\end{table}

We then need to form strongly-lensed pairs of events and unlensed pairs using the waveforms in our data sets. We know from the theory how a lensed wave should look like. We can express the amplitude $h_L$ of a strongly-lensed gravitational wave in the geometrical optics limit with
\begin{equation}
    h_L(\omega) = \sum_{j=1}^N |\mu_j|^{1/2} h(\omega) e^{-i\omega \Delta t_j - i \text{sign}(\omega) n_j \pi} \:,
\end{equation}
where $h(\omega)$ is the amplitude of the unlensed wave, $N$ is the number of images, $\mu_j$ and $\Delta t_j$ are the magnification and time delay of the $j$th image respectively. The Morse index $n_j$ is $0$, $1/2$ and $1$ for type I, II and III images respectively. The quantities $N$, $\mu_j$, $\Delta t_j$ and $n_j$ depend on the lens model that is considered and can be computed for a given lensing system.

To mimic lensing, we generate multiple events from each waveform, \textit{i.e.} project the waveform onto a network of detectors with random and different detection time, noise realisation and network SNR. We consider three detectors, Virgo~\cite{VIRGO_det} and LIGO Hanford and Livingston~\cite{LIGO_det}. For each waveform, we thus get a set of events that have the same instrinsic parameters, since they are generated from the same waveform, but different time of arrival and amplitude (random SNR and change in antenna pattern). The difference in arrival time mimics the lensing-induced time delay, while the different SNRs mimic different magnifications $\mu_j$ for each event. However, we do not use any lens model to generate those time delay and magnification, since they are generated randomly.

To form unlensed pairs, we can simply pair two events that were generated from two different waveforms. We generate as many lensed pairs as unlensed ones. Note also that we attribute an image type, hence a Morse phase, randomly to each gravitational wave. Therefore, a wave in an unlensed pair may still have a non-zero Morse phase. This corresponds to the case where one or two waves in the unlensed pair are independently lensed and thus come from different sources.

Although our choices of no lens model and same proportion of lensed and unlensed events do not correspond to realistic cases, it avoids biases in the network performance as it is trained to distinguish lensed pairs from unlensed ones solely based on the waveforms and independently of any model. We can thus evaluate the ability of a neural network to identify waveforms that are similar. Better results may be achieved by taking into account the lensing probability and adapting the proportion of lensed events accordingly, but the choices made in this work are sufficient for our purpose.\\

\begin{figure*}
    \centering
    \scalebox{0.85}{
    \begin{tikzpicture}

        \foreach \x in {5,4,...,0}
        {
            \node[rec,rotate=-90,fill=b,draw=white] at ({-5.3+\x * 0.1},{-0.5+\x * 0.1}) {$6\times1024 \, \in \mathbb{C}$};
        }

        \draw[link] (-3.8, 0.) -- (-2.,0.) node[midway, anchor=south, yshift=0.2cm] {\parbox{2.cm}{convolution}} node[midway, anchor=north, yshift=-0.2cm] {\parbox{2.cm}{+ \textbb{C}ReLU\\+ average pool}};

        \foreach \x in {5,4,...,0}
        {
            \node[rec,rotate=-90,fill=b,draw=white,opacity={1-\x * 0.18}] at ({-1+\x * 0.1},{-0.5+\x * 0.1}) {$64\times256 \, \in \mathbb{C}$};
        }

        \draw[link] (0.5, 0.) -- (3.5,0.) node[midway, anchor=south, yshift=0.2cm] {\parbox{2.5cm}{3$\times$ \\ convolution \\(channels $\times$ 2)}} node[midway, anchor=north, yshift=-0.2cm] {\parbox{2.8cm}{+ \textbb{C}ReLU \\+ average pool}};
        
        \foreach \x in {5,4,...,0}
        {
            \node[rec,rotate=-90,fill=b,draw=white,opacity={1-\x * 0.18}] at ({4.5+\x * 0.1},{-0.5+\x * 0.1}) {$512\times4 \, \in \mathbb{C}$};
        }

        \node at (0, 2.5) {\large{Convolution}};

        \node at (0., -4) {\large{(a)}};
    \end{tikzpicture}}
    
    \vspace{0.5cm}
    
    \scalebox{0.85}{
    \begin{tikzpicture}
        \foreach \x in {5,4,...,0}
        {
            \node[rec,rotate=-90,fill=b,draw=white] at ({-4+\x * 0.1},{-0.5+\x * 0.1}) {$6 \times 1024 \, \in \mathbb{C}$};
        }

        \draw[link] (-2.5, 0.) -- (-0.5,0.) node[midway, anchor=south, yshift=0.2cm] {Convolution};

        \foreach \x in {5,4,...,0}
        {
            \node[rec,rotate=-90,fill=b,draw=white,opacity={1-\x * 0.18}] at ({0.5+\x * 0.1},{-0.5+\x * 0.1}) {$512 \times 4 \in \mathbb{C}$};
        }
        
        \draw[link] (2, 0.) -- (4,0.) node[midway, anchor=north, yshift=-0.2cm] {\parbox{2.0cm}{flattening \\ + \\ concatenation}};
        
        \node[rec, minimum width=1.5cm, text width=1cm,fill=b,draw=white] at (3, 2.5) {$\Delta t$};

        \draw[thick] (3, 1.7) -- (3.,0.);
        
        \node[rec,rotate=-90,fill=b,draw=white, minimum width=6cm] at (5,0.){$4096+1 \in \mathbb{R}$};

        \foreach \x in {-1.75,-1.25,...,1.75}
        {
            \draw[fill=b] (7, {\x}) circle (0.2);
            \foreach \y in {-2.5,...,2.5} 
            {
                \draw[opacity=0.3,gray,thin] (5.6, \y) -- (6.8, \x);
            }
        }
        \node[rotate=-90] at (7,3) {\large{256}};
        \foreach \x in {-0.75,-0.25,0.25,0.75}
        {
            \draw[fill=b] (8., {\x}) circle (0.2);
            \foreach \y in {-1.75,-1.25,...,1.75} 
            {
                \draw[opacity=0.4,gray,thin] (7.2, \y) -- (7.8, \x);
            }
        }
        \node[rotate=-90] at (8,3) {\large{128}};
        \foreach \x in {-0.25,0.25}
        {
            \draw[fill=b] (9, {\x}) circle (0.2);
            \foreach \y in {-0.75,-0.25,0.25,0.75}
            {
                \draw[opacity=0.5,gray,thin] (8.2, \y) -- (8.8, \x);
            }
        }
        \node[rotate=-90] at (9,3) {\large{64}};
        \draw[fill=b] (10., 0) circle (0.2);
        \foreach \y in {-0.25,0.25}
        {
            \draw[opacity=0.6,gray,thin] (9.2, \y) -- (9.8, 0);
        }
        \node[rotate=-90] at (10,3) {\large{1}};
        \draw[link] (10.5, 0.) -- (11.5,0.) node[midway, anchor=south, yshift=0.2cm] {\large{$\sigma$}} node [pos=1,anchor=west,xshift=0.1cm] {\large{$y$}};

        \node at (4., -4) {\large{(b)}};
    \end{tikzpicture}}
        
    \caption{Illustration of the neural network (a) the convolution part of the CNN (b) the full architecture. The input consists of $0.5$s (at $2048$Hz) the strain 2 events, \textit{i.e.} 3 time series for each event corresponding to the strain of measured in the 3 detectors. These 6 time series are then stacked on top of each other, resulting in an input of shape $6\times 1024$. It is transformed into a complex value by setting the imaginary part to 0. The quantity $\Delta t$ is the time delay between the detection of the 2 events in the pair and $\sigma$ is the sigmoid activation function.}
    \label{fig:full-nn}
\end{figure*}
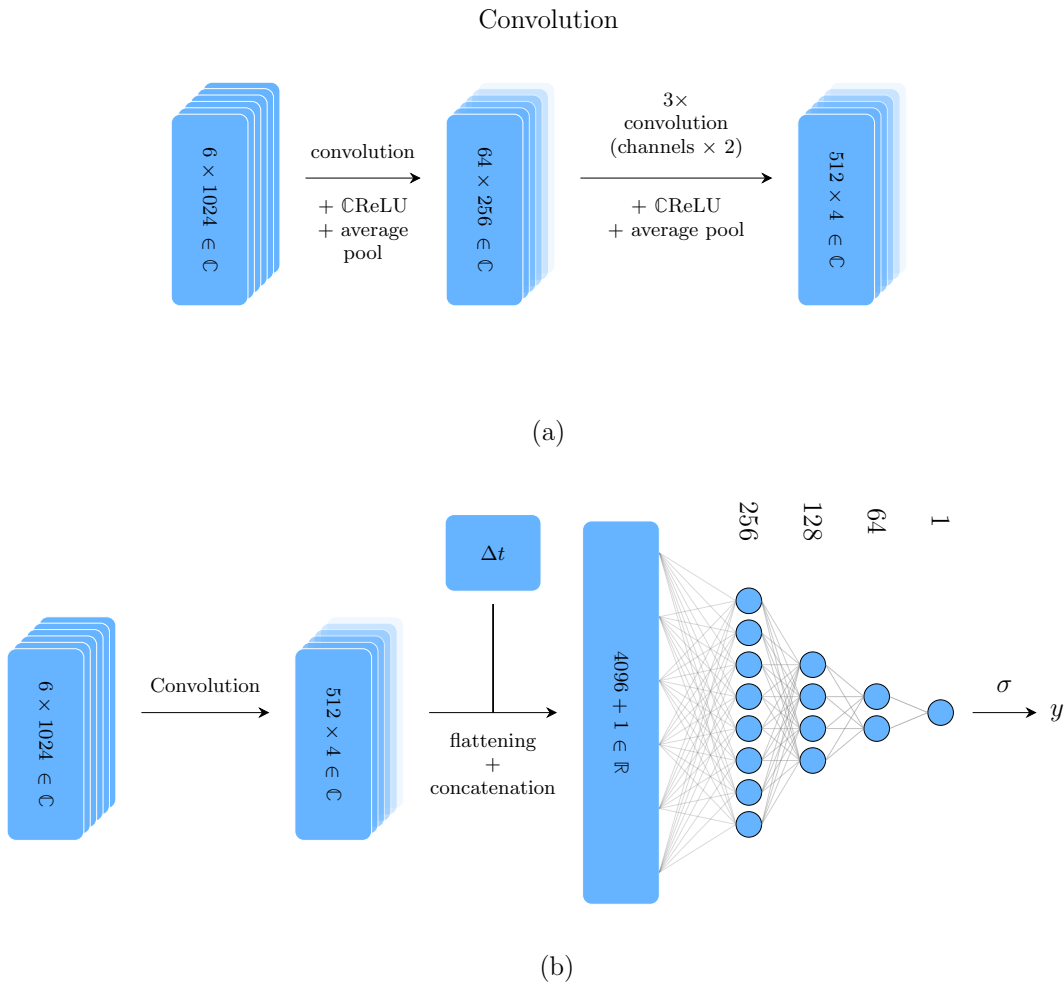

Finally, to allow for larger networks (deeper or with more neurons) and shorter time of evaluation, we truncate the time series to a duration of $0.5$s, which is long enough for BBH signals, and use a sampling frequency of $2048$Hz. We also whiten the final time series using O3 simulated PSDs provided by LALSuite~\cite{LAL} (XLALSimNoisePSDaLIGO140MpcT1800545 for LIGO detectors and XLALSimNoisePSDAdVO3LowT1800545 for Virgo) and apply a highpass filter at $20$Hz. 

In order to have a fixed reference time for all the events, we choose the $0.5$s of signal such that we have $0.42$s before the geocentric merger time and $0.08$s after. In other words, we take the strain of each detector in $t\in [t_c-0.42s,t_c+0.08s]$. This procedure thus aligns all the events with each other, with their respective geocentric merger time located at $0.42$s in the time series, but does not align the strains of each detector. If they were aligned, then the information of the time-delay between detectors, and hence of the sky position of the source, would be lost and could not be exploited by the network. Note, however, that no information about the sky position, such as skymaps, is given explicitly to the network and that there is no guarantee that it can fully use the information from the time delay between detectors.

In practice, the merger time is not known exactly. To make the model more realistic, we consider an error in the merger time and use the estimated merger time $\Tilde{t}_c$ rather than the true merger time to align the events, which results in a misalignment between the events. We could also align the waves at the merger time measured in one of the detector, which should have a lower uncertainty than the geocentric time. The results we present should also be valid for that alignment.

\subsection{\label{sec:nn_model}Neural network model}

We consider a convolutional neural network (CNN, \textit{e.g.}~\cite{CNN} and references therein), as it is easily implemented and a straightforward model for the task of identifying features. Since a main difference of time series with time-frequency maps is the phase, we decided to use complex-valued neural networks, \textit{i.e.} neural networks with complex weights. The network was implemented using \texttt{PyTorch}~\cite{Paszke_2019}.

\subsubsection{\label{sec:cv-nn}Complex-valued neural networks}

Complex-valued neural networks offer more choices regarding activation functions compared to real-valued ones. For example, one could use the usual activation function on the modulus and on the phase separately, or on the real and complex parts, again separately. In our case, we only use the \textbb{C}ReLU, which is defined as~\cite{Trabelsi_2018}
\begin{equation}
    \text{\textbb{C}ReLU}(\mathbf{z}) = \text{ReLU}(\text{Re}\{\mathbf{z}\}) + i \, \text{ReLU}(\text{Im}\{\mathbf{z}\})
\end{equation}

The same is naturally applied to the average-pooling layers, where we simply take the average of $\mathbf{z}$ over the kernel.\\

Another important difference that needs to be addressed is the weight initialization. The usual uniform or Gaussian He~\cite{He_2015} and Glorot~\cite{Glorot_2010} initialization showed, in some cases, some problems during the training where the network would not learn properly. Following~\cite{Trabelsi_2018}, we used a Rayleigh distribution to initialize the amplitude of the weights with the $\sigma$ parameter chosen to satisfy the variance required for either He or Glorot initialization. The phase of the complex numbers is drawn uniformly. 

\subsubsection{\label{sec:cnn}Convolutional neural network}

For the CNN, we use 4 complex-valued convolution layer with a kernel of size 3, followed by \textbb{C}ReLU activation and an average pooling of kernel 4. The first layer outputs $64$ channels, while the following convolutions each double the number of channels of their input. Then, the real and imaginary parts are concatenated along with the time delay between the events. After that, the network contains 4 real-valued fully connected layers with ReLU activation, except for the last layer. The output has a size of $1$ and we finally apply the sigmoid activation $\sigma$ to get the prediction of the neural network between $0$ and $1$. The architecture is illustrated in Fig.~\ref{fig:full-nn}. The input of the network consists of the time series of $2$ events, hence $6$ time series of $0.5$s each sampled at $2048$Hz. They are then stacked to get the final input of size $6\times 1024$ and transformed into complex numbers by setting the imaginary part to $0$. The first event in the input is always the one that was detected first.

\subsection{\label{sec:training}Training}

For training, we use a set of $10^5$ waveforms. Each waveform is used to generate 3 events. We can form 3 different pairs from these 3 events and, since they were generated from the same initial waveform, they will then correspond to the lensed pairs, as explained in section~\ref{sec:Data}. We also use events from different waveforms to form unlensed pairs. At each epoch, we only use $3$ unlensed pairs for each waveform, the second event of the pair being chosen randomly among all the other events. We limit this number of unlensed pairs to keep a balance between the number of lensed and unlensed pairs seen at each epoch of training. However, since the unlensed pairs are chosen randomly, they are different at each epoch. Also, at each epoch, the SNR of each event is drawn uniformly in $[7, 30]$ and the noise associated to the event is generated.

This technique virtually increases the size of our data set. The network will indeed see lensed pairs at different phases, since we use $3$ different pairs for a single waveform, and also see events in different noise and SNRs at each epoch. Although we indeed observed a reduction of the overfitting and validation loss, the effect is limited since the initial waveforms are still the same.

To ensure the robustness of our network on larger misalignment, we use a maximum error on the merger time of $20$ms. We therefore use the estimated merger time $\Tilde{t}_c$ for alignment, which is drawn randomly in $[t_c- 20\text{ms}, t_c + 20\text{ms}]$, where $t_c$ is the exact coalescence time, resulting in a maximum total misalignment of $40$ms between the two events in the pair.\\

\section{Results}

We will now discuss the performance of the network on our own data set and compare with~\cite{Goyal_2021} based on their data set. We compare the performance using the ROC curves that shows the true Positive Rate (TPR), \textit{i.e.} the proportion of lensed pairs that are identified, against the false Positive Rate (FPR), \textit{i.e.} the proportion of unlensed pairs that wrongly classified as being lensed. A good classifier has a high TPR at a low FPR.

Since the data sets contain only waveforms, and not events, we create the events at test time. This involves several randomly drawn parameters, namely the sky position, polarization, merger time (hence time delays), image type, merger time uncertainty (hence misalignment) and SNR. To account for those random parameters in the data, we repeat our tests multiple times and we will show the average results and related $1\sigma$ uncertainties.\\

\begin{figure}[h!]
    \centering
    \includegraphics[width=\linewidth]{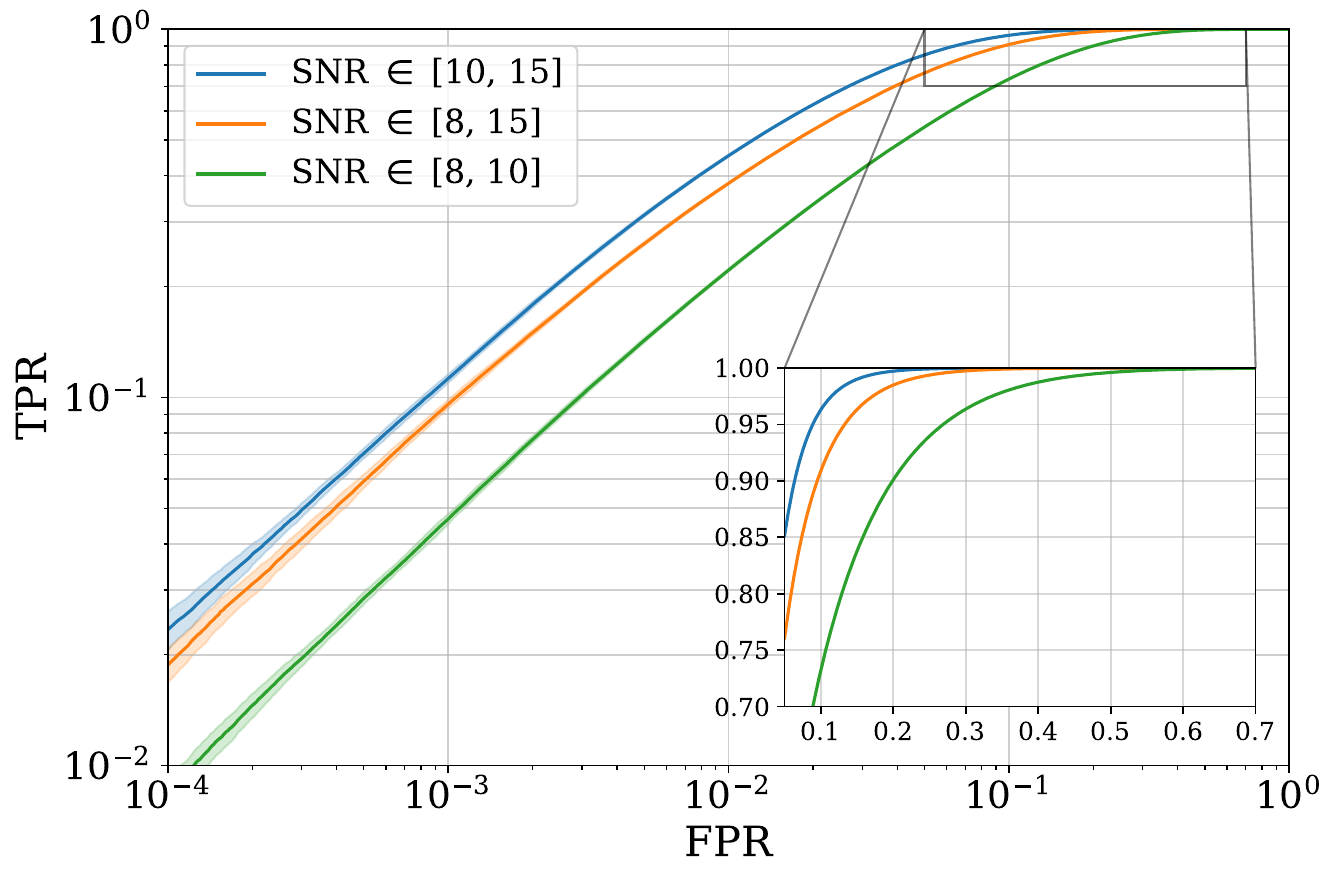}
    \caption{Performance of the CNN on our data set for different network SNR intervals, namely $[10,15]$ (blue), $[8,15]$ (orange) and $[8,10]$ (green). The shaded areas represent the $1\sigma$ uncertainty arising from the random parameters in the data. We can see that the SNR has a strong impact on the SNR.}
    \label{fig:roc_cnn}
\end{figure}

Our test set is composed of $5\cdot 10^4$ waveforms, each used to make $5$ events as explained in section~\ref{sec:Data}. We thus use $5\cdot 10^5$ lensed and unlensed pairs. We repeat the test $30$ times and show the resulting ROC curve on our data set in Fig.~\ref{fig:roc_cnn} for three different SNR intervals, with all three types of images, and a maximum error on the merger time of $1$ms. The shaded areas correspond to the $1\sigma$ uncertainty. We can see that the blue curve, corresponding to higher SNRs (between 10 and 15 for each event) is much higher than the green curve, corresponding to the lower SNRs (between 8 and 10 for each event). The maximum difference in TPR indeed reaches $\sim 0.3$ at an FPR of $\sim 0.04$. This indicates that the SNR has a strong impact on the performance of the network, as can be expected. We can also see that the network seems to have properly learned how to recognize similar waveforms at all SNRs, as the curves lie significantly above the diagonal.

There are however several factors that could potentially alter the performance of the network. As mentioned, we rely on the merger time to align the two events and consider some uncertainty on its measurement that would result in the misalignment. In the time-frequency representation, the time resolution is lower and this uncertainty may be contained within a single pixel. In our case, though, the misalignment is resolved and could then potentially cause trouble. Another possible source of error when applying our model on true data is that we use a waveform approximant and the true gravitational wave may not look exactly like the model. Although the error is expected to be small, it is worthwhile to check that the network is not sensitive to these differences. Finally, our model is also theoretically sensitive to the phase shift induced by lensing. This phase difference could then also potentially alter the performance of the network by either helping it finding lensed events, or preventing it from properly comparing the two waveforms.

\begin{figure}[h!]
    \centering
    \includegraphics[width=\linewidth]{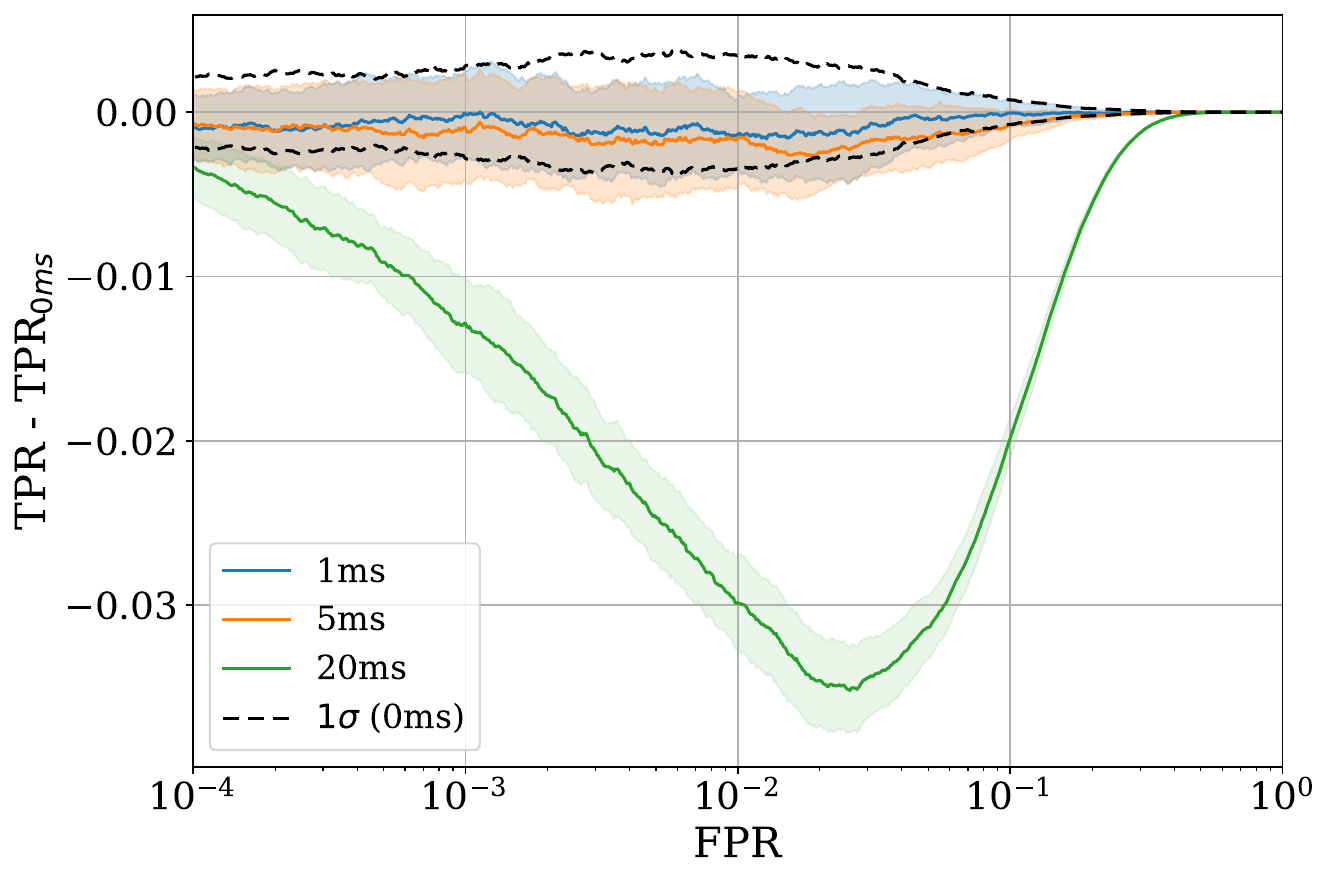}
    \caption{Effect of misalignment on the CNN performance for SNRs in $[8,15]$. We plot the difference in average TPR between the case of a maximum error on merger time of $1$ (blue), $5$ (orange) or $20$ms (green) and the case of exact alignment. The shaded areas represents the $1\sigma$ uncertainty, and the dashed line is the uncertainty for the reference case of exact merger time. The error remains close to $0$ and within the uncertainty of the reference for the $1$ and $5$ms cases. For larger misalignments, the difference is larger, but still much smaller than the effect of SNR.}
    \label{fig:comp_cnn_uc}
\end{figure}

We first analyse the effect of the misalignment of the two events in the pair on the performance of the network on our test set by considering different levels of errors on the merger time. In Fig.~\ref{fig:comp_cnn_uc}, we plot the difference of the average ROC curve between a case with misalignment and the reference case where the alignment is perfect, \textit{i.e.} with no error on the merger time. We consider a maximum error $\sigma_t = 1,5$ and $20$ms and, for each event, draw the estimated merger time (used for alignment) in $[t_c-\sigma_t, t_c+\sigma_t]$ where $t_c$ is the exact merger time. The maximum misalignement that is possible is thus $2\sigma_t$. As in the previous case, we repeat the test $30$ times and show the associated $1\sigma$ uncertainty. We can see that for a small error on the merger time, the effect of the misalignment is consistent with $0$, lying almost fully within the uncertainty of the $0$ms case. For the case of $5$ms, although on average it seems a bit worse than the $1$ms case, it is still also mostly contained within the  uncertainty of the reference case and very close to $0$. For a larger error, though, the effect is more pronounced with a maximum difference in TPR of $0.03$. It is however not strong enough to say that the network is not able to classify properly the pairs on average and is one order of magnitude smaller than the effect of low SNR. It could then be used on data with larger uncertainty on the merger time, \textit{i.e.} it does not require parameter estimation results to be run.

\begin{figure}[h!]
    \centering
    \includegraphics[width=\linewidth]{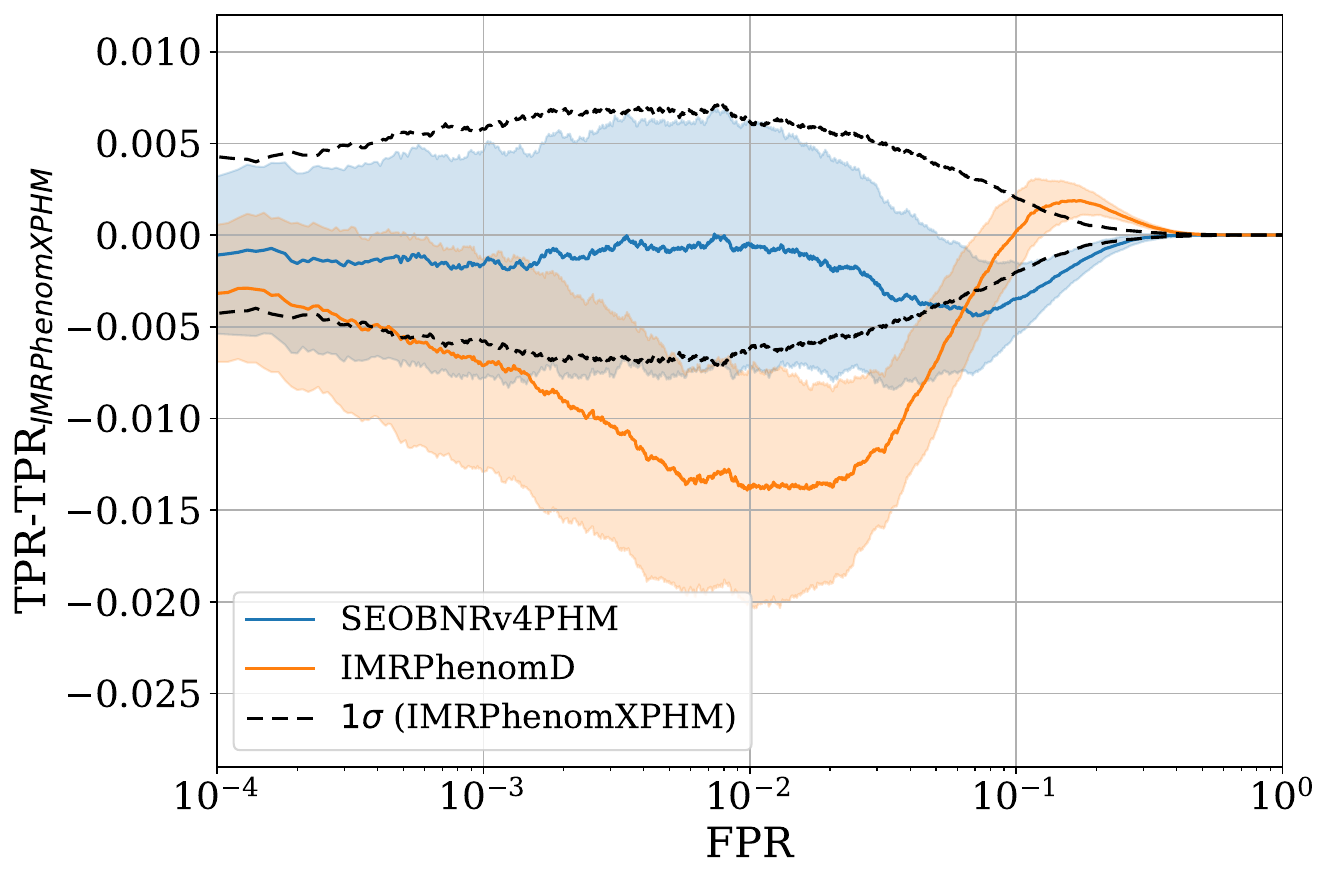}
    \caption{Comparison of the performance of the network using different waveform models. The curves represent the difference in TPR for \texttt{SEOBNRv4PHM} (blue) and \texttt{IMRPhenomD} (orange) with the one obtained with the reference model \texttt{IMRPhenomXPHM} used for training. The shaded areas correspond to the $1\sigma$ uncertainty. The difference is close to $0$ or to the uncertainty of the reference model. For both models, it does not change the global performance of the network.}
    \label{fig:comp_SEOB}
\end{figure}

Next, we check the performance of the network on other waveform models. For the comparison, we used the \texttt{SEOBNRv4PHM}~\cite{Ossokine_2020} waveforms as it based on a different approach than IMRPhenom, and the \texttt{IMRPhenomD}~\cite{Khan_2016,Husa_2016} waveforms that do not contain any higher order modes, in which case the Morse phase simply mimics a change in coalescence phase, even for type-II images~\cite{Ezquiaga_2021}. We generate a new test set by drawing $10^4$ samples of chirp masses, mass ratio, coalescence phase and inclination and generate a waveform with the 3 different approximants for each sample. Each waveform is then used to make $5$ events. We thus use $1\cdot 10^5$ lensed and unlensed pairs. Similarly to the other tests, we allow the sky localization, polarization, merger time, image type, misalignment and SNR to be different between different approximants. We use a maximum error on the merger time of $1$ms and SNRs in $[8,15]$. 
To account for all these random parameters, we perform $100$ independent tests and show the associated $1\sigma$ uncertainties in Fig.~\ref{fig:comp_SEOB}. We plot the difference in average ROC curves where we take the \texttt{IMRPhenomXPHM} as the reference since it was the one used in the training set. For the \texttt{SEOBNRv4PHM}, the difference remains mostly below $0.5\%$ at all FPRs, including uncertainties. It would thus only have an effect at very low FPRs, but it also lies within the uncertainties of the reference model in those region. The difference is therefore not significant. In the case of \texttt{IMRPhenomD}, the difference is more significant, but again an order of magnitude smaller than the maximum difference induced by low SNRs. This waveform does impact the performance at low FPR, but the global performance remains similar. This more significant difference could be due to the absence of higher modes that results in more significant differences with the reference waveform and also affects type-II images. To make the model more robust, we could include different waveform approximants in the training data, although it does not seem necessary given our results. \\

\begin{figure}[h!]
    \centering
    \includegraphics[width=\linewidth]{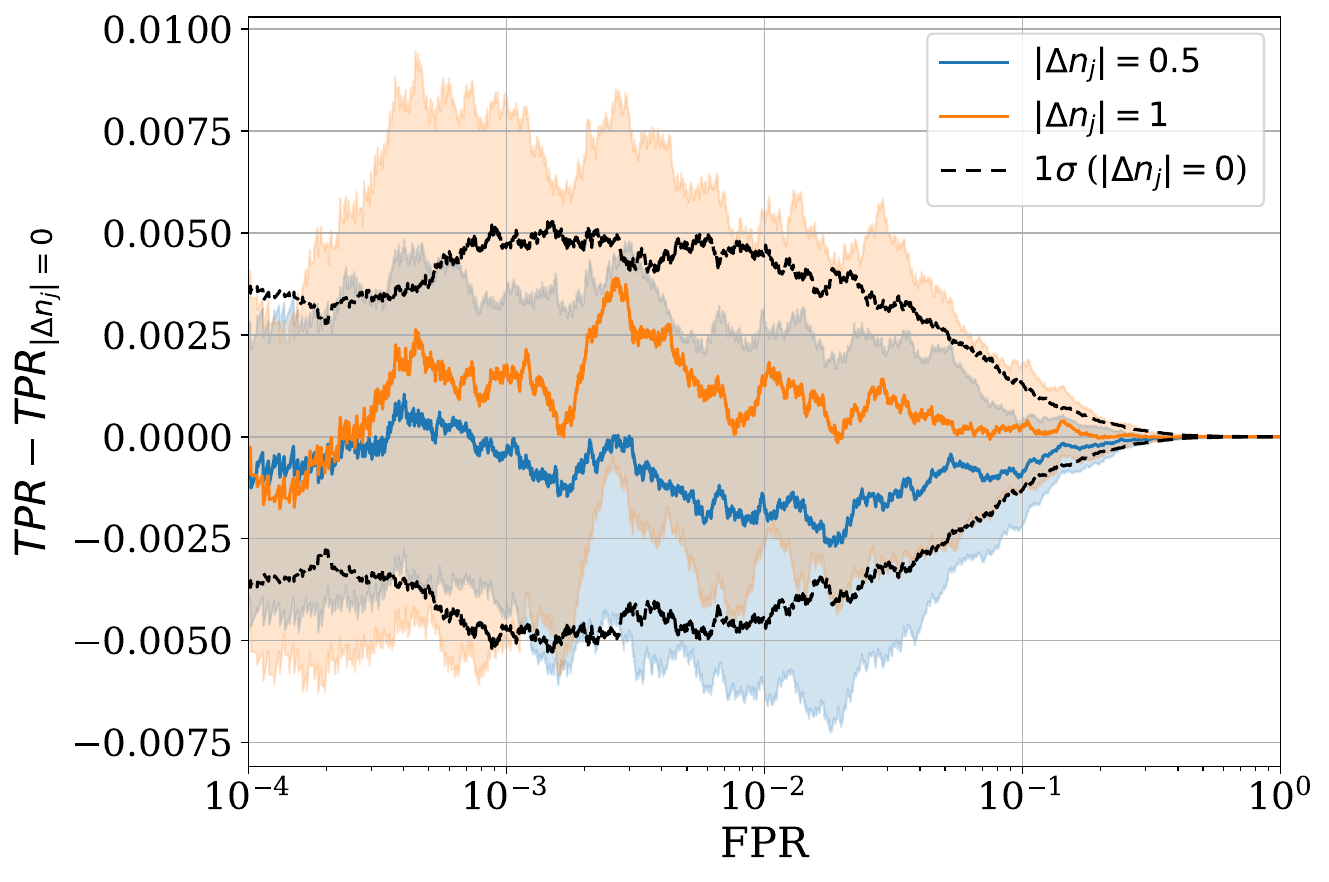}
    \caption{Comparison of the performance of the network for different Morse index differences. The curves correspond to the difference in TPR for pairs with a phase difference of $\pi/2$ (blue) or $\pi$ (orange) and pairs with no phase difference. The shaded areas correspond to the $1\sigma$ uncertainty. We can see that the curves all lie close to 0 and within the uncertainties of the reference case. The difference in Morse phases do therefore not impact the performance.}
    \label{fig:comp_morse}
\end{figure}

The last effect specific to the time domain representation we investigate is the effect of the phase shift (Morse phase). We use our test set with SNRs in $[8,15]$ and a maximum error on the merger time of $1$ms. We repeat the test $30$ times and compute the difference in the average ROC curves for different phase shifts compared to the reference case of no phase difference. We show the results in Fig.~\ref{fig:comp_morse}. We can see that both cases of phase difference between the waves lead to no significant effect on the performance, as the averages lie very close to $0$ and within the uncertainty of the reference case. It therefore appears that the phase shift does not prevent the network from identifying similar waveforms.\\

We will now compare our model to a model based on the time-frequency representation of the data. We compare our model to the one considered in~\cite{Goyal_2021}, referred to as G+21 in the following, since they use a model with no other addition than the time-frequency maps, contrary to Ref.~\cite{Magare_2024} where an additional representation is used. 

In their work, G+21 also presented a model where the skymaps of both events in the pair are used as input, but we compare our results to the model that does not use these.  
The reason for our choice of comparison is that we want to investigate the role of the representation of the strain data itself, without any additional analysis performed on it. We thus compare the performance of the models as single-step analyses, whereas generating skymaps would add an extra analysis. It would nevertheless be interesting to compare the performance of the two models when skymaps are included in the input for both representations, which is left for future work.

For the comparison, we generated events with the same parameters as the data set used in G+21, with the \texttt{IMRPhenomXPHM} approximant and with type-I images only. Contrary to our own data set, everything is fixed except for the estimated merger time that is used to align the events. Note that this new data set is very different from the one we used. First, there are many more unlensed pairs than lensed ones, as is expected. The data set contains $\sim 300$ lensed pairs and about $1000$ unrelated events used to make $\sim 5\cdot 10^5$ unlensed pairs. Then, the data in G+21 are generated taking into account a singular isothermal ellipsoid model for the lens, which makes the relation between two lensed waves stronger than our model since the magnification and time-delays are consistent with an underlying lens model, while they are random variables in our case. Finally, the regions of the parameter space spanned by the two data sets overlap but are not the same. The results may thus be more sensitive to overfitting.

\begin{figure}
    \centering
    \includegraphics[width=\linewidth]{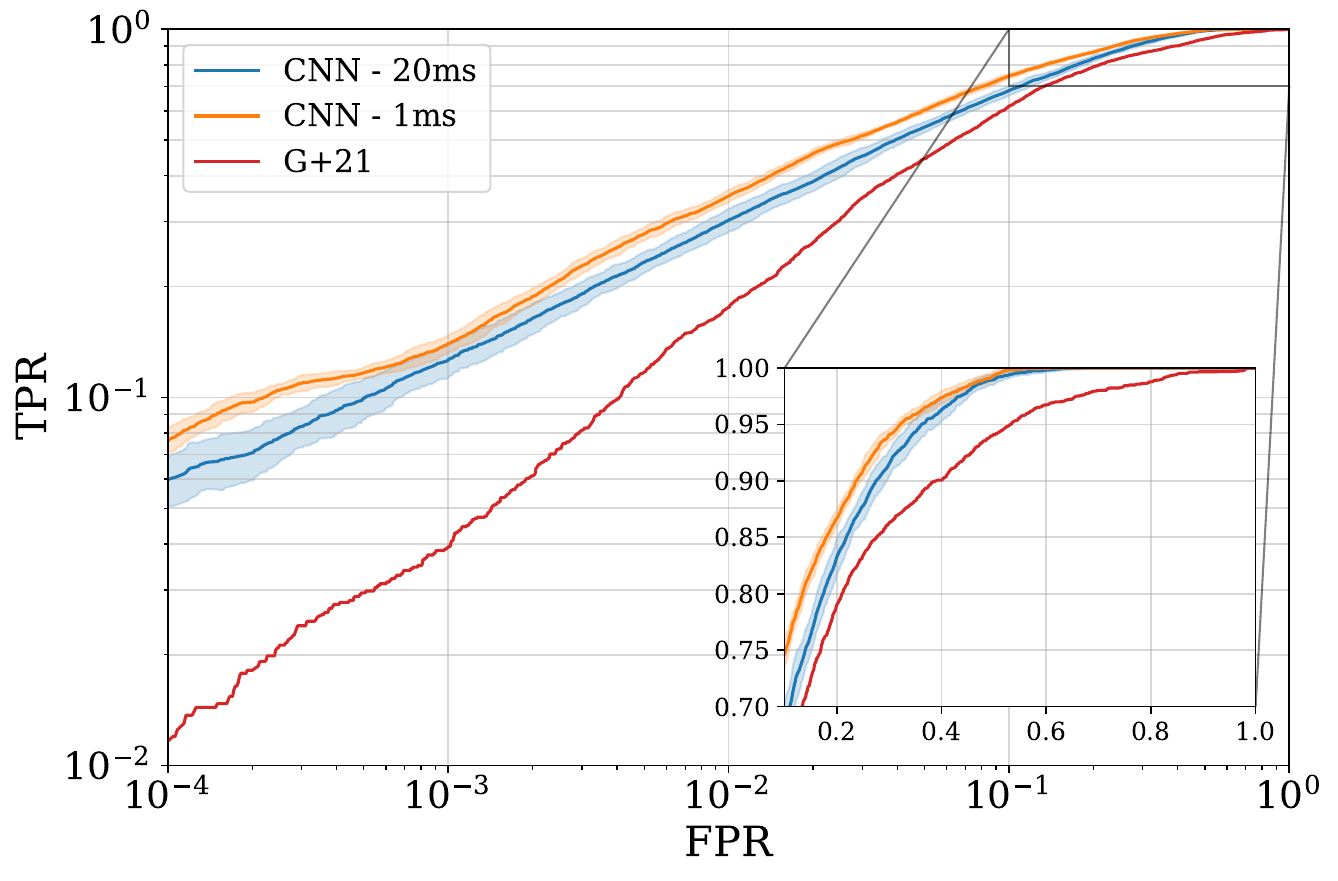}
    \caption{Comparison with the time-frequency model of G+21 (red) on their data set. The time in the legend refers to the uncertainty on the merger time for our model (blue and orange). The shaded areas represent the resulting $1\sigma$ uncertainty. The TPR of our model is above the TPR of G+21 at all FPRs, indicating that our model performs better.}
    \label{fig:comp_G21}
\end{figure}

The comparison on the data set from G+21 is shown in Fig.~\ref{fig:comp_G21}. We show the ROC curve of $20$ independent tests with a maximum error on the merger time of $1$ and $20$ms, resulting in the $1\sigma$ uncertainty represented by the shaded areas. Note that the uncertainty over multiple models showed in G+21 is not shown here, but the conclusion remains the same. We can clearly see that the the time series model performs better at all FPRs and that a larger misalignment does not change this conclusion. The difference peaks at an FPR of about $0.01$ where it is around $0.14$ and $0.2$ for the blue and orange curves, respectively. The difference is also very significant at the lowest FPR, where the TPR is $\sim 5$ times larger for our model.

It is important to stress that, although our model performs better, it is not straightforward to assert that it is due to the choice of representation. However, we use a similar architecture and a network with a lower number of parameters than the \textit{DenseNet201}~\cite{Huang_2017} used in G+21. It is therefore reasonable to assume that at least part of the reason is the choice of the representation. We also note that, because of the smaller size of the time series, we are able to consider larger networks for a same computation time. We could therefore consider bigger network and get better performance. This is an indirect advantage of the time series representation that could also partially explain our results.
Another advantage of our representation that could explain the results is the fact that our alignment procedure does not modify the time delay between detectors in the input, hence retaining information about the sky position as already mentioned.\\

Finally, we report the computation wall-time of our network on a GPU and CPU. On an NVIDIA GeForce GTX 750 GPU and for a batch of $500$ pairs, we measure a computation time of $8.4 \pm 0.97$ms, while it took $2.73 \pm 0.24$s on a CPU. In both cases, we can analyse a pair in less than a second. These times do not include the pre-processing of the time series, which is limited, in the case of the analysis of real events, to a whitening of the data, applying a highpass filter and cropping the time series to $0.5$s. Therefore, we expect that the pre-processing time, even if it is longer than the prediction time, would still allow for an analysis of less than a second per pair on a GPU. 

\section{Conclusion}

In this work, we trained a neural network to identify strongly-lensed gravitational-waves pairs based on the time series signal, as opposed to the time-frequency representation used in existing works. We motivate this choice by the presence of the phase that is otherwise loss in the transformation to a time-frequency plane. Also, we limit the pre-processing time using the time domain representation and keep the input one-dimensional, allowing for faster inference or larger networks.

We did not consider any lens model or realistic population in order to evaluate the ability of our neural network to identify lensed events solely based on the waveform, and hence show the potential of the time domain representation without any bias.

We tested the network on events containing Gaussian noise with a network SNR between $8$ and $15$ and with different image types. We align the waves based on their merger time but consider an uncertainty in the estimated merger time resulting in a misalignment of the events in the input. We show that the our model is able to properly identify lensed pairs, especially when the SNR is higher. The network was also tested on the data set used in~\cite{Goyal_2021} and shows better performance than their network that uses time-frequency maps. The two representations may however have different limitations, so that a combination of the two might result in even better performance. 

We analyzed the sensitivity of the network to the misalignment, to the waveform model and to the phase shift. We show that the misalignment does have a negative effect on the performance, but that effect is negligible compared to the effect of the SNR. The waveform model and the phase shift do not seem to have a significant effect either.\\

To improve our model further, we can train and test the model on real noise and add skymaps to the input. Indeed, even though the network has access to and is able to use some sky-position information through the time delay between detectors, we have no guarantee that this information is fully used. Including skymaps in the input would give explicitly the sky position and should make it easier for the network to find unlensed events solely based on the different sky positions. Another extension of the work could include the difference in phase in the training strategy. For example, the network could be trained to distinguish $4$ different case: unlensed, unlensed but possible presence of a phase shift, lensed and lensed with possible presence of phase shift. This would indeed be theoretically possible using the time series, since the phase is retained, and it would also allow the network to spot additional lensing events.

Finally, it would be interesting to train and test the network on data using a lens model and using a more realistic population. We could then compare the performance with this work and the one from~\cite{Magare_2024}. This would allow us to assert better the influence of the data representation (time domain and time-frequency domain) and of the models that are used. These ideas are left for future work.

\begin{acknowledgments}

We would like to thank Jean-René Cudell for starting the project and for his help at the earlier stages of development. We also thank Srashti Goyal for sharing her data. We are grateful for computational resources provided by the LIGO laboratory and supported by the National Science Foundation. The code makes use of the  \texttt{PyTorch}~\cite{Paszke_2019}, \texttt{NumPy}~\cite{Harris_2020}, \texttt{Bilby}~\cite{Ashton_2019} and \texttt{scikit-learn}~\cite{pedregosa_2011} packages. This work was partially supported by grants from The Research Foundation Flanders (FWO) under project(s) G086722N.

\end{acknowledgments}

\bibliography{bibliography}{}

\end{document}